\documentclass{amsart}

\usepackage[cp1251]{inputenc}
\usepackage[T2A]{fontenc}
\usepackage{array}
\usepackage{amssymb}
\usepackage{amsmath}
\usepackage{amsfonts}
\usepackage{amsthm}
\usepackage{latexsym}
\usepackage{indentfirst}
\usepackage{bm}
\usepackage{enumerate}
\usepackage{graphicx}
\usepackage{epsf}
\usepackage{wrapfig}
\usepackage{euscript}
\usepackage{indentfirst}
\usepackage[english]{babel}

\newcommand*{\mailto}[1]{\href{mailto:#1}{\nolinkurl{#1}}}

\begin{document}

\title[Interest rate models and Whittaker functions]{Interest rate models and Whittaker functions}

\author[D.\ Muravey]{Dmitry Muravey}
\address{GEOLAB LLC \\ Ordzhonikidze 12\\ Moscow \\ Russia}
\email{d.muravey@mail.ru}
\maketitle

\begin{abstract}
I present the technique which can analyse some interest rate models: Constantinides-Ingersoll, CIR-model, geometric CIR and Geometric Brownian Motion. All these models have the unified structure of Whittaker function. The main focus of this text is closed-form solutions of the zero-coupon bond value in these models. In text I emphasize the specific details of mathematical methods of their determination such as Laplace transform and hypergeometric functions.
\end{abstract}

\section{Introduction and Summary}

\par
The motivation of this paper is results analogical to the affine class of models. The main advantage of affine theory is the explicit form of solution for complete ensemble of the models including Vasicek, Ho-Lee, Hull-White(so-called extended Vasicek model) and Cox- Ingersoll-Ross model. Whereas exponential affine representation is not applicaple to Constantinides-Ingersoll, Geometric CIR, Dothan, Brennan and Schwartz, Black-Karasinski and etc. However for a subset of these models we can obtain the unified structure that can be described in terms of Whittaker equation. This is not a rare equation in financial mathematics, for example the Asian option value can be expressed in terms of it \cite{MUR}. First introduction of this ODE was in \cite{WH}. Application of this equation has come from physics, it is the Schrodinger equation with Morse potential and it describes the hydrogen atom. In mathematical field it is  limiting form of Rhiemman's differential equation\cite{AS}. Also this equation can be described with the theory of eigenfunctions expansions \cite{T}. A solution of it is the Whittaker function which has closer relations with other special functions such as confluent hypergeometric function, Kummer function, Bessel function, Error function, Parabolic cylinder function and etc. In spite of very complicated analytical structure (Whittaker function does not have a unified series or integral representation) it has known asymptotic, analytical continuation laws and a lot of relations to less complicated functions. Basing on these facts I use the unified technology which is similar in all investigating models. The steps are the following: At first I use Feynman-Kac formula for PDE boundary problem determination, then I make Laplace transform, solve corresponded non-homogeneous ODE boundary problem and make inversion.
\par
This approach is demonstrated on the zero coupon bond value with maturity date \begin{math} T \end{math}, that can be defined as this expectation:
\begin{equation}
    P(r,t)=E\bigg[\exp{ \big\{-\int\limits_{t}^{T} r(\xi)d\xi \big\}} \bigg].
\end{equation}
and interest rate \begin{math} r(t) \end{math} is modeled by SDE:
\begin{equation}
\begin{split}
   &\quad Constantinides -Ingersoll \quad model. \quad  dr(t)=\alpha r^2 (t) dt +\beta r^{3/2}(t) dW_t,
   \\
   &\quad Geometric \quad Brownian \quad motion. \quad dr(t)=\alpha r(t) dt +\beta r(t) dW_t,
   \\
   &\quad Geometric  \quad CIR  \quad process. \quad dr(t)= (\alpha r^2(t) +\gamma r(t)) dt +\beta r^{3/2}(t) dW_t,
   \\
   &\quad Original \quad CIR \quad process. \quad dr(t)= (\alpha r(t) +\gamma ) dt +\beta r^{1/2}(t) dW_t,
\end{split}
\end{equation}
where \begin{math} W_t \end{math} is a standard Wiener process.
\par
After using Feynman-Kac formula we obtain the boundary problem to the parabolic PDE with variable but time-independent coefficients. Contrary to the constant coefficients (e.g. Black-Scholles equation) the construction of fundamental solution(also so called Green function) is not so obvious and in general case it has no closed-form expression. However if coefficients have special polynomial structure it is possible to construct Green function. We can follow scheme: in initial condition we set Dirac delta function, make Laplace transform, solve corresponding ODE and make inversion. The obtained functional is the Green function of the problem and therefore the solution of the original problem can be represented as the convolution product with initial function. Hence the main question is how to solve the corresponded ODE. In general cases it has no solution in closed-form, but in models 1 and 2 corresponded ODE can be reduced to the Bessel equation that it the special case of Whittaker equation and in models 3 and 4 to the original Whittaker equation \cite{AS},\cite{NU}. For formulas simplification the boundary problem is solved directly without the fundamental solution construction. In this way there are several technically sophisticated moments: the nonhomogeneous ODE solving and inversion of Laplace transform which includes the branch points and residuals under Bessel and Whittaker functions.
\par
The paper follows this structure: Sections 2, 3, 4 and 5 are about Constantinides-Ingersoll, Geometric Brownian motion, geometric CIR and original CIR models respectively. Each section contains PDE problem definition and direct/inverse Laplace transform description. The appendix A contains a one of approaches for the hypergeometric equation. In Appendix B I outline main steps of Laplace transform method in the time-depended affine models (Ho-Lee and Hull-White). And for comfortable reading the Appendix C contains the necessary facts about Bessel and Whittaker functions.

\section{Constantinides -Ingersoll model}
\subsection{PDE definition}
This model was introduced in \cite{CI} and it including the presence of taxes. The bond value \begin{math} P(r,t) = P(r,\tau) \end{math} where \begin{math} T-t=\tau \end{math} is a solution of this PDE boundary problem
    \begin{equation}
    \left\{ {\begin{array}{l}
        (\beta^2 r^3 /2)  P_{rr} + \alpha r^2 P_r -rP =P\tau, \\
        P(r,0)=1, \\
        P(0,\tau)=1, \\
        P(+\infty,\tau)=0. \\
    \end{array}} \right.
    \end{equation}

\subsection{Laplace transform and connection to the Bessel equation}
\par
In the beginning a new variable is denoted as:
\begin{equation}
    y=1/\sqrt{r}, \quad
    P(r,\tau)=Y(y,\tau), \quad
    P_r =-\frac{-Y_y}{2r\sqrt{r}}, \quad
    P_{rr}=\frac{Y_{yy}}{4r^3}+\frac{3Y_{y}}{4r^2\sqrt{r}}.
\end{equation}
The equation (3) is transformed to
\begin{equation}
    \frac{\beta^2}{8}Y{yy} +\frac{3\beta^2-4\alpha}{8y}Y_{y}- \frac{1}{y^2}Y=Y_\tau.
\end{equation}
The representation \begin{math} Y(y,\tau)=1+y^{2\lambda+1}W(y,\tau)  \end{math} is used to make zero boundary and zero initial conditions
\begin{equation}
\left\{ {\begin{array}{l}
    W_{yy} +W_y /y -4\mu^2W /y^2 = 8W_\tau /{\beta^2}  + 8y^{(-3-2\lambda)}/\beta^2,  \\
    W(y,0)=W(0,\tau)=W(+\infty,\tau)=0. \\
\end{array}} \right.
\end{equation}
Constants in equation (6) are defined by
\begin{equation}
\lambda= -1 +\alpha/\beta^2, \quad \mu=\sqrt{1/4+2/\beta^2+\alpha^2/\beta^4-\alpha/\beta^2}, \quad \sigma=2\sqrt{2}/\beta.
\end{equation}
Direct and inverse Laplace transform are defined by formulas (8), the last equality is a notation of operation calculus (the relation between source and image of transformation)
\begin{equation}
\begin{split}
    V(y, \eta)= \int\limits_{0}^{+\infty} {W(y,\tau)  e^{-\eta \tau } d\tau}& , \quad
    W(y,\tau)=\frac{1}{2\pi i} \int\limits_{N-i\infty}^{N+i\infty} {V(y, \eta) e^{\eta \tau } d\eta}, \\
    &W(y, \tau) \fallingdotseq V(y,\eta).
\end{split}
\end{equation}
The image of Laplace transform is this nonhomogeneous ODE with zero boundary conditions:
\begin{equation}
\left\{ {\begin{array}{l}
    V'' +V'/y -4\mu^2 V / y^2 - \sigma^2 \eta V  = \sigma^2 y^{(-3-2\lambda)}/\eta, \\
    V(0)=V(+\infty)=0. \\
    \end{array}} \right.
\end{equation}
Change of variable \begin{math} \xi=\sigma y\sqrt{\eta} \end{math}  transforms equation (9) to the nonhomogeneous Bessel equation \cite{AS}
\begin{equation}
\left\{ {\begin{array}{l}
    V'' +V' / \xi -(1+4\mu^2 / \xi^2) V =\sigma^{3+2\lambda}\eta^{\lambda-1/2}  \xi^{-3-2\lambda} , \\
    V(0)=V(+\infty)=0. \\
    \end{array}} \right.
\end{equation}
The solution of homogeneous equation is linear combination of modified Bessel functions \begin{math} K_{2\mu}(\xi) \end{math} and \begin{math} I_{2\mu} (\xi) \end{math} (for more details see Appendix C)) and the solution of problem (10) is given by representation
\begin{equation}
    V(\xi)=\sigma^{3+2\lambda} \eta^{\lambda-1/2} \bigg( K_{2\mu}(\xi) \int\limits_{0}^{\xi} \phi^{-2(1+\lambda)} I_{2\mu}(\phi)  d\phi + I_{2\mu}(\xi) \int\limits_{\xi}^{+\infty} \phi^{-2(1+\lambda) }K_{2\mu}(\phi) d\phi \bigg),
\end{equation}
Formally the solution of equation (10) is represented by the sum of homogeneous solution and solution with right-hand side function. It is easy to show that homogeneous solution branch equals zero(It is based on modified Bessel functions asymptotic, see Appendix C.)

\subsection{Laplace transform inversion: residuals and branch points}
\par
Let's consider in detail the inversion procedure that is defined by formula (8)
\begin{equation}
\begin{split}
    W(y,\tau) &=\frac{1}{2\pi i} \int\limits_{N-i\infty}^{N+i\infty}
    \bigg( \frac{\sigma^{3+2\lambda}}{\eta^{1/2-\lambda}} K_{2\mu}(\sigma \sqrt{\eta}y)
    \int\limits_{0}^{\sigma y \sqrt{\eta}}  \phi^{-2(1+\lambda)}  I_{2\mu}(\phi) d\phi +
     \\ &+\frac{\sigma^{3+2\lambda}}{\eta^{1/2-\lambda}}  I_{2\mu}(\sigma\sqrt{\eta}y)
     \int\limits_{\sigma y \sqrt{\eta} }^{+\infty} \phi^{-2(1+\lambda)}  K_{2\mu}(\phi) d\phi  \bigg) e^{\eta \tau} d\eta.
\end{split}
\end{equation}
The variable \begin{math} \xi \end{math} can also be used as integration variable. It transposes the path of integrate \begin{math} (N-i\infty, N+i\infty) \end{math} to the curve \begin{math} L_{\xi} \end{math} (see Figure 1).
\begin{figure}
\begin{center}
    \resizebox*{12cm}{!}{\includegraphics{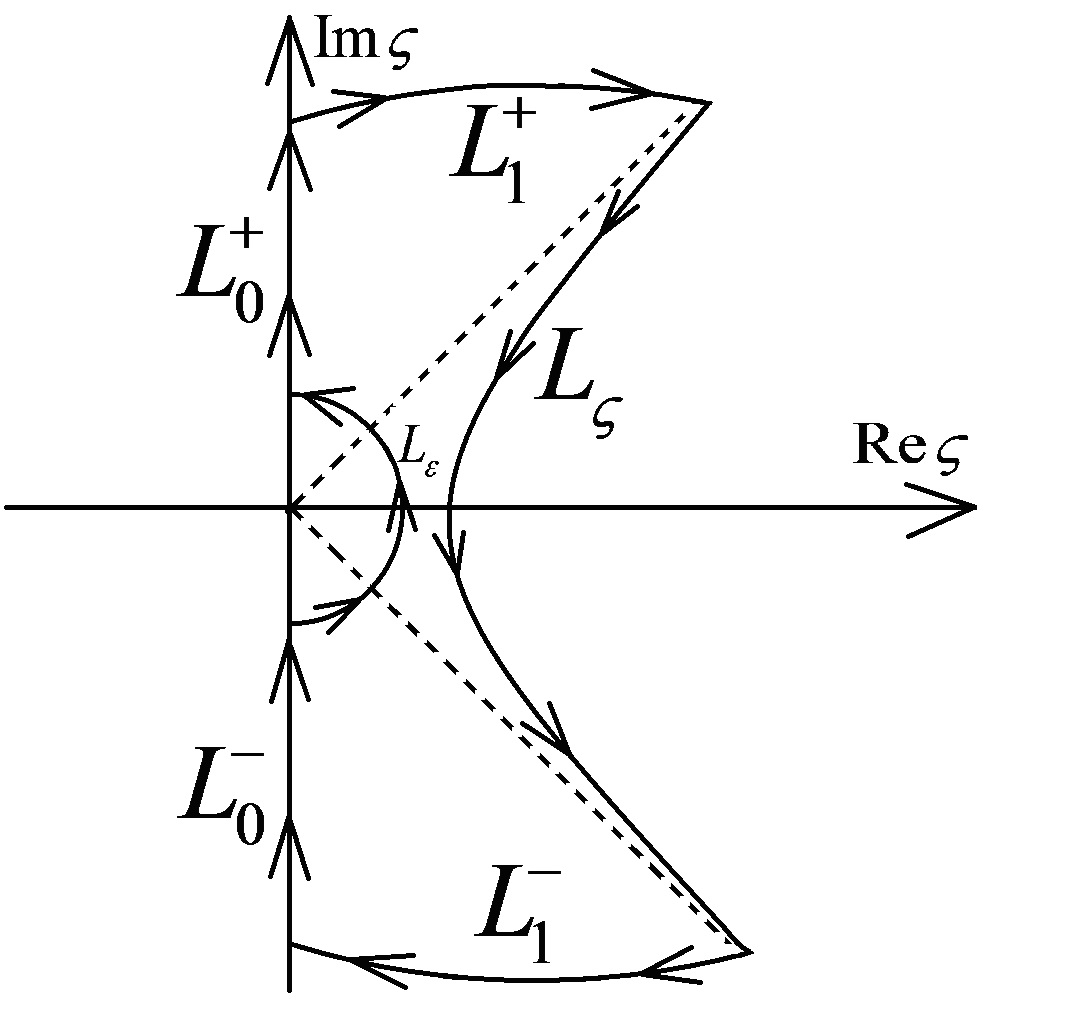}}
    \caption{\label{fig1} Integration loop in Constantinides-Ingersoll model
    \label{sample-figure}}
\end{center}
\end{figure}

\begin{equation}
W(y,\tau) =\frac{1}{2\pi i} \int\limits_{L_{\xi}} \frac{2\sigma^2 y^{-2\lambda-1} \Psi(\xi)}{\xi^2} \exp \bigg\{ \frac{\xi^2 \tau}{\sigma^2 y^2} \bigg\}  d\xi.
\end{equation}
where
\begin{equation}
\Psi(\xi) = \xi^{2(\lambda+1)} \bigg( K_{\mu}(\xi) \int\limits_{0}^{\xi}  \phi^{-2(1+\lambda)}  I_{2\mu}(\phi) d\phi + I_{2\mu}(\xi)
     \int\limits_{\xi}^{+\infty} \phi^{-2(1+\lambda)}  K_{2\mu}(\phi) d\phi  \bigg).
\end{equation}
Complex integral on the Loop \begin{math}  L= L_\xi+ L_1^+ + L_1^- + L_2^+ + L_2^- + L_\epsilon  \end{math} equals zero:
\begin{equation}
\int\limits_{L} \frac{2\sigma^2 y^{-2\lambda-1} \Psi(\xi)}{\xi^2} \exp \bigg\{ \frac{\xi^2 \tau}{\sigma^2 y^2} \bigg\}  d\xi=0.
\end{equation}
Integral on the path \begin{math} L_2^+ + L_2^- \end{math} converges to zero under Jordan's lemma. Integral on the \begin{math} L_\epsilon \end{math} yields half of the residue in the zero point which equals
\begin{equation}
\frac{1}{2} Res \bigg(\frac{2\sigma^2 y^{-2\lambda-1} \Psi(\xi)}{\xi} \exp \bigg\{ \frac{\xi^2 \tau}{\sigma^2 y^2} \bigg\}, \xi=0 \bigg) =\sigma^2 y^{-2\lambda-1} \lim_{\xi \rightarrow 0} \frac{\Psi(\xi)}{\xi} = y^{-2\lambda-1}.
\end{equation}
The residue determination is based on modified Bessel functions asymptotic(see Appendix C) and L'Hopital rule. Therefore function \begin{math} W(y,\tau) \end{math} has this form
\begin{equation}
 W(y,\tau)= -\frac{1}{2\pi i } \int\limits_{L_0^{-} + L_0^{+}} \frac{2\sigma^2 y^{-2\lambda-1} \Psi(\xi)}{\xi^2} \exp \bigg\{ \frac{\xi^2 \tau}{\sigma^2 y^2} \bigg\} d\xi - y^{-2\lambda-1}.
 \end{equation}
On the path \begin{math} L_{0}^{-} \end{math} variable \begin{math} \xi=-i \theta \end{math} and on \begin{math} L_{0}^{+} \end{math} is set \begin{math} \xi=i \theta  \end{math}. This yields
\begin{equation}
 W(y,\tau)= \frac{\sigma^2 y^{-2\lambda-1}}{\pi} \int\limits_{0}^{+\infty} \frac{\Psi(i\theta) +\Psi(-i\theta)}{\theta^2} \exp \bigg\{ \frac{-\theta^2 \tau}{\sigma^2 y^2} \bigg\} d\theta - y^{-2\lambda-1}.
\end{equation}
The integrand function \begin{math} \Psi(\zeta) \end{math}  has this property (Proof is based on analytical continuation laws, see Appendix C)
\begin{equation}
\Psi(e^{\pi i /2}\theta) +\Psi(e^{-\pi i /2}\theta)= \pi  \theta^{2(\lambda+1)}  J_{2\mu} (\theta) \int\limits_0^{+\infty} \phi^{-2(1+\lambda)} J_{2\mu} (\phi) d\phi,
\end{equation}
where \begin{math} J_{\nu}(z) \end{math} is the Bessel function \cite{AS}. Then using the well-known integral \cite{GR}:
\begin{equation}
\int\limits_{0}^{+\infty} x^{\mu} J_{\nu} (ax) dx = 2^{\mu} a^{-\mu-1} \frac{\Gamma(1/2+\nu/2+\mu/2)}{\Gamma(1/2+\nu/2-\mu/2)}, \quad -\Re\nu-1 < \mu < 1/2, \quad a>0.
\end{equation}
Return to original variables gives this closed-form expression of bond value \begin{math} P(r,t) \end{math}
\begin{equation}
    P(r,t)=\frac{\Gamma(1/2+\mu-\lambda)}{\Gamma(1/2+\mu+\lambda)} \int\limits_{0}^{+\infty} (\theta/2)^{2\lambda} e^{-\beta^2 \theta^2 r(T-t)/8} J_{2\mu} (\theta) d\theta.
\end{equation}
Note that bond value depends only on the multiplication of rate \begin{math} r \end{math} and time \begin{math} T-t \end{math}.
\subsection{The hypergeometric equation and the Whittaker equation}
For connection to the hypergeometric equation the variable \begin{math} x=r \tau \end{math} is used directly in problem (3). This operation transforms PDE boundary problem to the ODE boundary problem
\begin{equation}
\left\{ {\begin{array}{l}
    (\beta^2 x^2/2)  F_{xx} + ( \alpha x -1 \big) F_x -F =0, \\
    F(0)=1, \\
    F(+\infty,t)=0.
\end{array}} \right.
 \end{equation}
\par
For solution determination the structure of equation coefficients is essential. It is the second order polynomial at the second order derivative, the linear polynomial at the first order derivative and the constant at function - this is a hypergeometric type ODE\cite{NU} by definition. Using advantages of hypergeometric theory(see Appendix A) function \begin{math} F \end{math} can be presented as
\begin{equation}
    F(x)=x^{1-\alpha/\beta^2} \exp{ \bigg( -\frac{1}{\beta^2 x} \bigg)} G \Bigg( \frac{2}{\beta^2 x} \Bigg),
\end{equation}
where \begin{math} G(z) \end{math} is a solution of Whittaker equation \cite{AS, NU}.
\begin{equation}
 G''(z)+G(z)\bigg(-\frac{1}{4}+\frac{\lambda}{z}+\frac{1/4-\mu^2}{z^2} \bigg)=0,
\end{equation}
constants \begin{math} \lambda \end{math} and \begin{math} \mu \end{math} are defined by (7).
\par Two linearly independent solutions of this equation are the Whittaker functions of the first kind \begin{math} M_{\lambda,\mu}(z) \end{math} and \begin{math} M_{-\lambda,\mu}(z) \end{math}. Let me mention that for special cases \begin{math} 2\mu= \pm 1, \pm 2, \pm 3,... \end{math} Whittaker functions of the first kind have singularities. In this case the Whittaker functions of the second kind \begin{math} W_{\lambda,\mu}(z) \end{math} and \begin{math} W_{\lambda,-\mu}(-z) \end{math} are used that are linear combinations of \begin{math} M_{\lambda,\mu}(z) \end{math} and \begin{math} M_{-\lambda,\mu}(z) \end{math}(see Appendix C). Therefore function \begin{math} F \end{math} is
\begin{equation}
    F(x)=C_1 x^{-\lambda} \exp \bigg(-\frac{1}{\beta^2 x} \bigg) M_{\lambda,\mu} \bigg(\frac{2}{\beta^2 x} \bigg) + C_2 x^{-\lambda} \exp \bigg(-\frac{1}{\beta^2 x} \bigg) M_{\lambda,-\mu} \bigg(\frac{2}{\beta^2 x}  \bigg).
\end{equation}
Asymptotic of Whittaker functions(see Appendix C) determines constants \begin{math} C_1 \end{math} and \begin{math} C_2 \end{math} and function \begin{math} F(x) \end{math} is expressed as
\begin{equation}
    F(x)= \frac{\Gamma(1/2+\mu -\lambda )}{\Gamma (1+ 2\mu)} \bigg(\frac {2}{\beta^2 x} \bigg)^{\lambda}  \exp \bigg(-\frac{1}{\beta^2 x} \bigg) M_{\lambda,\mu} \bigg(\frac{2}{\beta^2 x} \bigg).
\end{equation}
And in original variables the bond value \begin{math} P(r,t) \end{math} have this second representation
     \begin{equation}
    P(r,t)= \frac{\Gamma(1/2+\mu -\lambda/2 )}{\Gamma (1+ 2\mu)} \bigg(\frac {2}{\beta^2 r (T-t)} \bigg)^{\lambda}  \exp \bigg(-\frac{1}{\beta^2 r (T-t)} \bigg) M_{\lambda,\mu} \bigg(\frac{2}{\beta^2 r (T-t)} \bigg).
    \end{equation}
Let me mention that formulas (21) and (27) are similar \cite{GR}
\begin{equation}
\begin{split}
     \int\limits_{0}^{+\infty} x^{\mu} e^{-\alpha x^2} J_{\nu}(\beta x) dx =\frac{\Gamma(\nu/2+\mu/2 +1/2 )}{\beta \alpha^{\mu/2}\Gamma (\nu+1)} e^{-\beta^2/8\alpha} M_{\mu/2,\nu/2} \big(\beta^2 /4\alpha \big),\\
     \quad \Re \alpha>0, \quad \Re(\nu+\mu)>-1,  \quad \beta>0.
\end{split}
\end{equation}
\subsection{CIR VR Model}
CIR VR model was introduced by \cite{COX} and focused on variable rate securities. It is the special case of Constantinides - Ingersoll model with the following restrictions
\begin{equation}
    \alpha=0, \quad \lambda = -1, \quad \mu =\sqrt{1/4+ 2/\beta^2}.
\end{equation}
The simplifications are substantial in both formulas (21) and (28)
\begin{equation}
    P_{VR}(r,t)=\frac{8}{\beta^2} \int\limits_{0}^{+\infty} \frac{J_{\sqrt{1+ 8/\beta^2}} (\theta)}{\theta^2} e^{-\beta^2 \theta^2 r(T-t)/8} d\theta,
\end{equation}

\begin{equation}
    P_{VR}(r,t)= \frac{\beta^2 r (T-t)\Gamma(3/2+\mu)}{2 \Gamma (1+ 2\mu)} \exp \bigg(-\frac{1}{\beta^2 r (T-t)} \bigg) M_{-1,\sqrt{1/4+2/\beta^2}} \bigg(\frac{2}{\beta^2 r (T-t)} \bigg).
\end{equation}

\section{Geometric Brownian Motion model}
\subsection{PDE definition and Laplace transform}
This model was considered by \cite{MAR}
For this model corresponded PDE defined as
    \begin{equation}
    \left\{ {\begin{array}{l}
        (\beta^2 r^2/2) P_{rr}+\alpha r P_r-rP =P_\tau, \\
        P(r,0)=1, \\
        P(0,\tau)=1, \\
        P(+\infty,\tau)=0. \\
    \end{array}} \right.
    \end{equation}
Contrary to the Black-Sholles equation this is impossible to use variable \begin{math} \ln r \end{math} that linearises equation.   However in terms of variable \begin{math} y=\sqrt r, \end{math} this model is analogical to the Constantinides - Ingersoll model. With the use Laplace transform \begin{math} P(r,\tau)=W(y,\tau) \fallingdotseq V(y,\eta) \end{math}  equation is transformed to the Bessel-type equation
\begin{equation}
    \left\{ {\begin{array}{l}
        V_{yy}+(4\lambda+3)V_{y}/y + V(-\sigma^2 -\sigma^2\eta/y^2)=-\sigma^2/y^2, \\
        V(0)=1/\eta, \\
        V(+\infty)=0. \\
    \end{array}} \right.
    \end{equation}
According to the previous section the solution of homogeneous equation is linear combination of modified Bessel functions
\begin{equation}
V(y) = C_1 y^{-2\lambda-1} K_{\nu}(\sigma y)+ C_2 y^{-2\lambda-1} I_{\nu}(\sigma y), \quad   \nu =\sqrt{(2\lambda+1)^2+\sigma^2 \eta}.
\end{equation}
Then the solution of problem (33) is expressed by
\begin{equation}
    V(y,\eta)=\sigma^2 y^{-2\lambda-1} \bigg(K_{\nu}(\sigma y) \int\limits_{0}^{y} \phi^{2\lambda} I_{\nu}(\sigma \phi) d\phi+ I_{\nu}(\sigma y) \int\limits_{y}^{+\infty} \phi^{2\lambda} K_{\nu}(\sigma \phi) d\phi  \bigg).
\end{equation}
or after change of integration variable
\begin{equation}
    V(y,\eta)=\sigma^2 \bigg(K_{\nu}(\sigma y) \int\limits_{0}^{1} \phi^{2\lambda} I_{\nu}(\sigma y \phi) d\phi+ I_{\nu}(\sigma y) \int\limits_{1}^{+\infty} \phi^{2\lambda} K_{\nu}(\sigma y \phi) d\phi  \bigg).
\end{equation}
The double integral can be reduced to the single integral. It is based on relation between modified Bessel functions and original Bessel functions\cite{GR}
\begin{equation}
\int\limits_{0}^{+\infty} xJ_{\nu}(ax) J_{\nu}(bx) \frac{dx}{x^2+c^2}=
    \left\{ {\begin{array}{l}
        I_{\nu}(bc)K_{\nu}(ac), \quad 0<b<a,\Re c>0,\Re \nu>-1,  \\
        I_{\nu}(ac)K_{\nu}(bc), \quad 0<a<b,\Re c>0,\Re \nu>-1,
    \end{array}} \right.
\end{equation}
and formula (21). Then the second representation of (36) is
\begin{equation}
V(y,\eta) = \sigma^2 2^{2\lambda}\frac{\Gamma(1/2+\lambda + \nu/2)}{\Gamma(1/2-\lambda + \nu/2)} \int\limits_{0}^{+\infty}  \frac{\zeta^{-2\lambda} J_{\nu} (\zeta) d \zeta}{\zeta^2+\sigma^2 y^2}.
\end{equation}
Due to the asymptotic of Bessel function (see Appendix C) this formula exists only for \begin{math} 2\lambda+1>-3/2 \end{math}. Let me also mention that in Constantinides - Ingersoll model the spectrum parameter \begin{math} \eta \end{math} is in argument of Bessel function, whereas in this case it is in the index.
\subsection{Inverse Laplace transform overview}
Inversion procedure have two sophisticated moments: firstly function \begin{math} V(y,\eta) \end{math} have residual in point \begin{math} \eta=0 \end{math}. Function \begin{math} V(y,\eta) \end{math} can have singularity (Integrals in (36) can be indefinite or Euler Gamma function in (38) has residuals in all non-positive integer points). Secondly the original function \begin{math} W(y,\tau) \end{math} do not have a unified representation (the representations (36) and (38) are both used).
\subsubsection{Residuals}
If \begin{math} 2\lambda+1 >0 \end{math} the function \begin{math} V(y,\eta) \end{math} does not have not residual. If \begin{math} 2\lambda+1 < 0 \end{math} function \begin{math} V(y,\eta) \end{math} has residual in the point \begin{math} 2\lambda+1+\nu =0 \end{math}. For its determination representation (39) is used.
\begin{equation}
Res \bigg(\Gamma(1/2+\lambda + \nu/2),\quad \eta=0 \bigg)= \frac{4(2\lambda +1) }{\sigma^2}.
\end{equation}
Therefore
\begin{equation}
Res \bigg( V(y,\eta)e^{\eta \tau},\quad \eta=0 \bigg)= \frac{2^{2\lambda+2}}{\Gamma(-2\lambda-1)} \int\limits_{0}^{+\infty}  \frac{\zeta^{-2\lambda} J_{-2\lambda-1} (\zeta) d \zeta}{\zeta^2+\sigma^2 y^2}.
\end{equation}
This integral is known\cite{GR}
\begin{equation}
\int\limits_{0}^{+\infty} x^{\nu+1} J_{\nu}(ax) \frac{dx}{x^2+b^2} =b^{\nu} K_{\nu} (ab) ,\quad a>0,b>0, -1<\Re\nu<3/2.
\end{equation}
Then the residual has this unified expression:
\begin{equation}
Res \bigg( V(y,\eta)e^{\eta \tau},\quad \eta=\eta_k \bigg)= \left\{ {\begin{array}{l}
        \frac{2}{\Gamma(-2\lambda-1)} \bigg(\frac{\sigma y}{2} \bigg)^{-2\lambda-1} K_{-2\lambda-1} (\sigma y),\quad 2\lambda+1<0, \\
        0, \quad 2\lambda+1>0.
            \end{array}} \right.
\end{equation}

\subsubsection{Laplace transform under branch points}
The integration loop \begin{math} L \end{math} is set analogical to Constantinides-Ingersoll model, see Figure 2. Path \begin{math} (N-i\infty,N+\infty) \end{math} is denoted as \begin{math} L_{\eta} \end{math}. The point \begin{math} \eta=-(2\lambda+1)^2/\sigma^2 \end{math} is the branch point. Then the function \begin{math} W(y,\tau) \end{math} can be represented
\begin{equation}
W(y,\tau) = -\frac{1}{2\pi i}\int\limits_{L^{+}+L^{-}} V(y,\eta) e^{\eta \tau} d\eta +Res \bigg( V(y,\eta)e^{\eta \tau},\quad \eta=0 \bigg)
\end{equation}
\begin{figure}
\begin{center}
    \resizebox*{12cm}{!}{\includegraphics{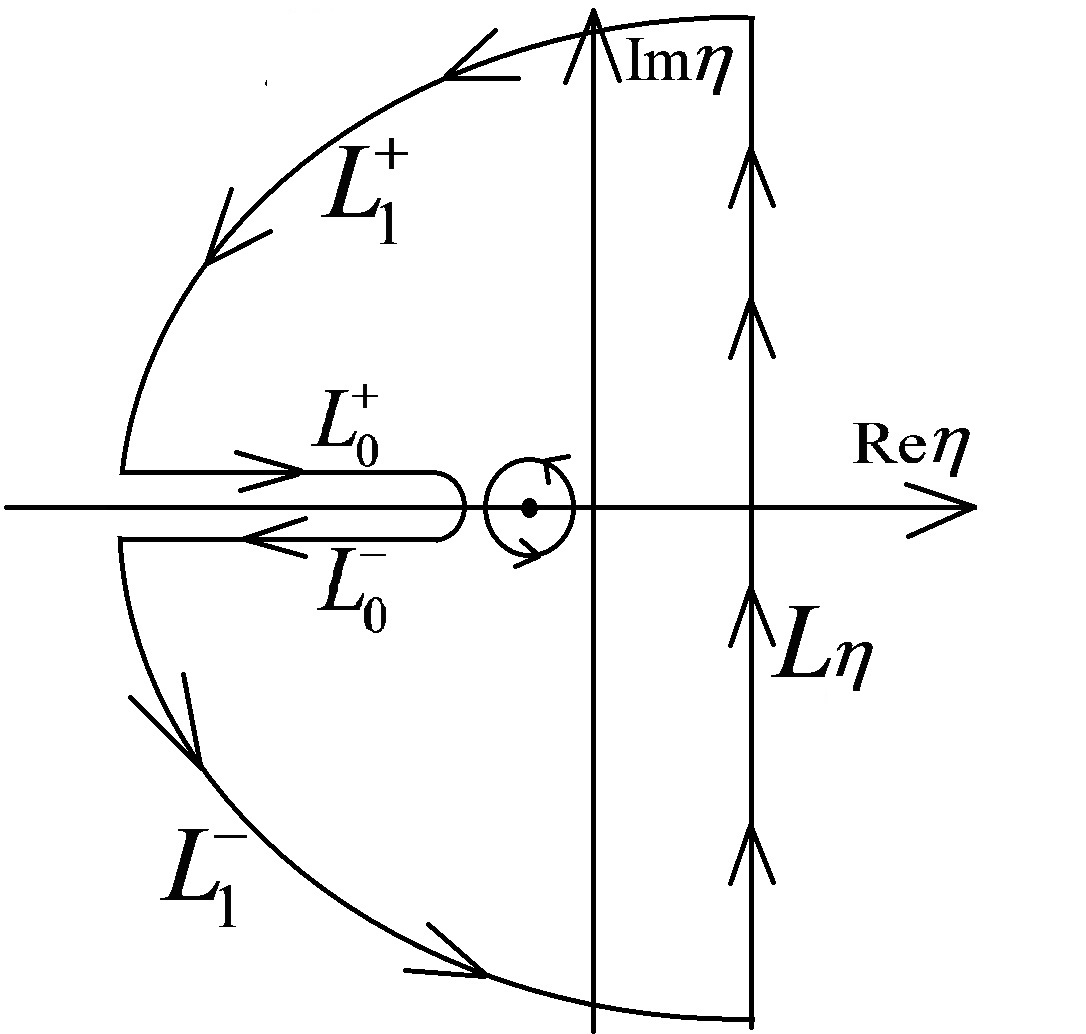}}
    \caption{\label{fig1} Integration loop in Geometric Brownian motion model
    \label{sample-figure}}
\end{center}
\end{figure}

\par Let's consider the case \begin{math} 2\lambda +1 > 0\end{math}. In representation (37) integration variable is set \begin{math} \sigma^2 \eta = -(2\lambda+1)^2 +e^{i\pi} \zeta \end{math} on path \begin{math} L_{0}^{+}\quad (\nu = i\sqrt{\zeta}) \end{math} and \begin{math} \eta = -(2\lambda+1)^2 +e^{-i\pi} \zeta \end{math} on \begin{math} L_{0}^{-} \quad (\nu = -i \sqrt{\zeta}) \end{math} (Also the property of modified Bessel function \begin{math} K_{\nu} (z) = K_{-\nu} (z) \end{math} is used). Then
\begin{equation}
\begin{split}
W(y,\tau) = -&\frac{e^{-\tau(2\lambda+1)^2/\sigma^2}}{2\pi i} \int\limits_{0}^{+\infty} e^{-\zeta \tau /\sigma^2 } \bigg(K_{i\sqrt{\zeta}}(\sigma y) \int\limits_{0}^{1} \phi^{2\lambda} \big(I_{-i\sqrt{\zeta}}(\sigma y \phi)-I_{i\sqrt{\zeta}}(\sigma y \phi) \big) d\phi +\\
 &+\big( I_{-i\sqrt{\zeta}}(\sigma y) -I_{i\sqrt{\zeta}}(\sigma y) \big) \int\limits_{1}^{+\infty} \phi^{2\lambda} K_{i\sqrt{\zeta}}(\sigma y \phi) d\phi  \bigg)d\zeta.
\end{split}
\end{equation}
The relation between functions \begin{math} K_{\nu}(z) \end{math} and \begin{math} I_{\nu}(z) \end{math} yields (see Appendix C)
\begin{equation}
W(y,\tau) = -\frac{e^{-\tau(2\lambda+1)^2/\sigma^2}}{2\pi^2 i} \int\limits_{0}^{+\infty} \int\limits_{0}^{+\infty} e^{-\zeta \tau /\sigma^2 } \phi^{2\lambda} \sin{\pi i \sqrt{\zeta}} K_{i\sqrt{\zeta}}(\sigma y) K_{i\sqrt{\zeta}}(\sigma \phi y) d\phi d\zeta.
\end{equation}
For reduction to the single integral this relation is used  \cite{GR}
\begin{equation}
\int\limits_{0}^{+\infty} x^{\mu} K_{\nu}(a x)dx = 2^{\mu-1} a^{-\mu-1}
\Gamma\bigg(\frac{1+\mu+\nu}{2} \bigg) \Gamma\bigg(\frac{1+\mu-\nu}{2} \bigg), \quad \Re(\mu+1\pm \nu )>0, \Re a>0.
\end{equation}
In original notations the bond value is represented by the formula
\begin{equation}
\begin{split}
&W(y,\tau) = \frac{e^{-(T-\tau)(2\lambda+1)^2/\sigma^2}}{2\pi^2 }  \bigg( \frac{\sigma\sqrt{r}}{2}\bigg)^{-2\lambda-1} \times  \\ \times \int\limits_{0}^{+\infty} \Gamma \bigg(\frac{1+2\lambda+i\theta}{2} \bigg) &\Gamma \bigg(\frac{1+2\lambda-i\theta}{2} \bigg) \theta e^{-\theta^2 (T-t) /\sigma^2 } \sinh{\theta \pi}  K_{i\theta}(\sigma \sqrt{r})d\theta.
\end{split}
\end{equation}
In the opposite case \begin{math} 2\lambda+1<0 \end{math} representation (38) should be used. This yields another representation
\begin{equation}
\begin{split}
P(r,t) &= \frac{2}{\Gamma(-2\lambda-1)} \bigg(\frac{\sigma \sqrt{r}}{2} \bigg)^{-2\lambda-1} K_{-2\lambda-1} (\sigma \sqrt{r})+ \frac{e^{-(T-\tau)(2\lambda+1)^2/\sigma^2}}{\pi i } \times\\
&\times \int\limits_{-\infty}^{+\infty}\int\limits_{0}^{+\infty} \frac{\Gamma(1/2+\lambda+i\theta/2)}{\Gamma(1/2-\lambda+i\theta/2)} \theta \bigg( \frac{\zeta}{2}\bigg)^{-2\lambda} \frac{e^{-(T-t)\theta^2/\sigma^2}J_{i\theta}(\zeta)}{\zeta^2+\sigma^2 r}  d\theta d\zeta.
\end{split}
\end{equation}

\section{The Geometric CIR model}
\subsection{PDE definition, Laplace transform and connection to the Whittaker equation}
This model about inflation rate factor was introduced in original Cox, Ingersoll, Ross paper in 1980. The corresponded PDE has this form
    \begin{equation}
    \left\{ {\begin{array}{l}
        (\beta^2 r^3 /2)  P_{rr} + (\alpha r^2+\gamma r) P_r -rP =P_{\tau}, \\
        P(0,\tau)=1, \\
        P(+\infty,\tau)=0. \\
        P(r,\tau)=1. \\
    \end{array}} \right.
    \end{equation}
The solution \begin{math} P(r,t) \end{math} is represented as the following structure and the Laplace transform is made (Note that this change of variable \begin{math} z \end{math} is not applicable to the Constantinides-Ingersoll model; this structure appears if we make zero coefficient of the first derivative).
\begin{equation}
P(r,t)= z^{\lambda}e^{z/2} Z(z,\tau), \quad z=\frac{2 \gamma }{\beta^2 r}, \quad \tau = T-t, \quad Z(z,\tau) \fallingdotseq G(z,\eta).
\end{equation}
Hence problem  (52) is reduced to the nonhomogeneous boundary problem for the Whittaker equation
\begin{equation}
    \left\{ {\begin{array}{l}
        G''+G\bigg(-1/4+\tilde{\lambda}/z+(1/4-\mu^2)/z^2 \bigg) =z^{-\lambda-1} e^{-z/2}/\gamma,\\
        e^{z/2} z^{\lambda} G \rightarrow 0,  \quad z\rightarrow 0, \\
        e^{z/2} z^{\lambda} G \rightarrow 2\gamma /\beta^2, \quad z\rightarrow +\infty.
    \end{array}} \right.
\end{equation}
where
\begin{equation}
    \tilde{\lambda} =-\lambda -\eta /\gamma, \quad \mu =\sqrt{1/4+2/\beta^2+\alpha^2/\beta^4-\alpha/\beta^2}.
\end{equation}
Contrary to the Constantinides - Ingersoll model linearly independent functions are Whittaker functions \begin{math} W_{\tilde{\lambda},\mu}(z) \end{math} and \begin{math} M_{\tilde{\lambda},\mu}(z) \end{math}. Using asymptotic and Wronskian of these functions(see Appendix C) the solution is
\begin{equation}
\begin{split}
G(z,\eta)=\frac{1}{\gamma} &\frac{\Gamma(1/2+\mu-\tilde{\lambda})}{\Gamma(1+2\mu)} \bigg( M_{\tilde{\lambda},\mu} (z) \int\limits_{0}^{z} \phi^{-\lambda-1} e^{-\phi/2} W_{\tilde{\lambda},\mu}(\phi)d\phi + \\
+& W_{\tilde{\lambda},\mu} (z) \int\limits_{z}^{+\infty} \phi^{-\lambda-1} e^{-\phi/2} M_{\tilde{\lambda},\mu}(\phi)d\phi \bigg).
\end{split}
\end{equation}
Usage of following relation between modified Bessel function and Whittaker functions (it is analogical to the formula (38))
\begin{equation}
\begin{split}
\int\limits_{0}^{+\infty} e^{-(a_1+a_2)t\cosh (x) /2 } \bigg[\coth \bigg(\frac{x}{2} \bigg) \bigg]^{2\nu} I_{2\mu} (t\sqrt{a_1a_2} \sinh{x}) dx = \\= \frac{\Gamma(1/2+\mu-\nu)}{t\sqrt{a_1a_2}\Gamma(1+2\mu)}  W_{\nu,\mu}(a_1 t) M_{\nu,\mu}(a_2 t), \quad
\Re (1/2+\mu-\nu)>0,\Re \mu>0, a_1>a_2.
\end{split}
\end{equation}
gives the double integral representation
\begin{equation}
G(z,\eta) = \frac{\sqrt{z}}{\gamma}  \int\limits_{0}^{+\infty} \int\limits_{0}^{+\infty} \phi^{-\lambda-1/2} e^{-\phi/2 -(z+\phi)\cosh{\zeta}/2} \big(\coth(\zeta/2) \big)^{2\tilde{\lambda}} I_{2\mu} (\sqrt{z \phi} \sinh {\zeta}) d\zeta d\phi.
\end{equation}
It can be reduced to the single integral expression due to this relation
\begin{equation}
\begin{split}
\int\limits_{0}^{+\infty} x^{\mu-1/2}e^{-\alpha x}  I_{2\nu} (2\beta \sqrt{x}) dx = \frac{\Gamma(\mu+\nu+1/2)}{\beta\Gamma(2\nu+1)}e^{-\beta^2/2\alpha} \alpha^{-\mu} M_{-\mu,\nu}(\beta^2/\alpha),  \\ \quad\Re(\mu+\nu+1/2)>0.
\end{split}
\end{equation}
Finally function \begin{math} G \end{math} represented as
\begin{equation}
G(z,\eta) = \frac{2e^{-z/2}}{\gamma} \frac{\Gamma(3/2-\alpha/\beta^2 + \mu)}{\Gamma(2\mu + 1)} \int\limits_{0}^{+\infty}  \frac{e^{-z\sinh^2{\zeta}/2} \coth^{2\tilde{\lambda}}\zeta }{\sinh{\zeta} \cosh^{1-2\lambda}\zeta} M_{\lambda, \mu} (z\sinh^2{\zeta}) d\zeta.
\end{equation}
Let's denote
\begin{equation}
\ln \tanh^2 \zeta =\psi, \quad \frac{2d\zeta}{\sinh \zeta \cosh \zeta }=d\psi, \quad \sinh^2 \zeta =\frac{e^{\psi}}{1-e^\psi}, \quad \cosh^2 \zeta=\frac{1}{1-e^\psi}.
\end{equation}
This yields
\begin{equation}
G(z,\eta) = \frac{e^{-z/2}}{\gamma}\frac{\Gamma(3/2-\alpha/\beta^2 + \mu)}{\Gamma(2\mu + 1)} \int\limits_{-\infty}^{0}  \frac{e^{-ze^{\psi}/2(1-e^\psi)} e^{-\psi\tilde{\lambda}}} {(1-e^{\psi})^{\lambda}} M_{\lambda, \mu} \bigg(\frac{ze^{\psi}}{1-e^\psi}\bigg) d\psi.
\end{equation}
Then the inverse Laplace transform is given by formula
\begin{equation}
\begin{split}
&Z(z,\tau)=\frac{e^{-z/2}\Gamma(1/2-\lambda + \mu)}{\gamma \Gamma(2\mu + 1)}\times\\
\times \frac{1}{2\pi i}  \int\limits_{N-i\infty}^{N+i\infty} \int\limits_{-\infty}^{0}
 &e^{\psi\eta/\gamma +\tau\eta} e^{-ze^{\psi}/2(1-e^\psi)}
 \bigg(\frac{e^{\psi}}{1-e^\psi}\bigg)^{\lambda} M_{\lambda, \mu} \bigg(\frac{ze^{\psi}}{1-e^\psi}\bigg) d\eta d\psi.
\end{split}
\end{equation}
Based on properties of Dirac delta function
\begin{equation}
\quad \frac{1}{2\pi i}  \int\limits_{N-i\infty}^{N+i\infty} e^{\psi\eta/\gamma +\tau\eta} d\eta = \delta(\psi/\gamma+\tau), \quad \int\limits_{-\infty}^{0}
 \delta(\psi/\gamma +\tau) H(\psi) d\psi =\gamma H(-\tau\gamma)
\end{equation}
inversion gives
\begin{equation}
Z(z,\tau)=\frac{e^{-z/2}\Gamma(1/2-\lambda + \mu)}{\Gamma(2\mu + 1)}
 e^{-ze^{-\tau\gamma}/2(1-e^{-\tau\gamma})}
 \bigg(\frac{e^{-\tau\gamma}}{1-e^{-\tau\gamma}}\bigg)^{\lambda} M_{\lambda, \mu} \bigg(\frac{ze^{-\tau\gamma}}{1-e^{-\tau\gamma}}\bigg).
\end{equation}
and the bond value \begin{math} P(r,t) \end{math} equal
\begin{equation}
    P(r,t)=\frac{\Gamma(1/2-\lambda + \mu)}{\Gamma(2\mu + 1)}
    \exp \bigg\{-\frac{1}{\beta^2 x(r,t)}  \bigg\}
    \bigg( \frac{2}{\beta^2 x(r,t)}\bigg)^{\lambda}
    M_{\lambda, \mu} \bigg( \frac{2}{\beta^2 x(r,t)} \bigg),
\end{equation}
where
\begin{equation}
x(r,t) = \frac{r(1-e^{-\gamma(T-t)})}{\gamma e^{-\gamma(T-t)}}, \quad \lim_{\gamma \rightarrow 0} x(r,t) =r(T-t).
\end{equation}
The bond value in the Constantinides - Ingersoll model can be obtained as the limiting form. Let me also mention than variable \begin{math} x(r,t) \end{math} can be used directly in equation (52). It also transforms PDE problem to the ODE problem in the manner of section 2.3.

\section{The Cox-Ingersoll-Ross Model}
\subsection{PDE definition and Laplace transform}
Original CIR model has this PDE boundary problem
    \begin{equation}
    \left\{ {\begin{array}{l}
        (\beta^2 r /2)  P_{rr} + \big( \alpha r + \beta \big)P_r -rP =P\tau, \\
        P(r,0)=1, \\
        P(0,\tau)=1, \\
        P(+\infty,\tau)=0. \\
    \end{array}} \right.
    \end{equation}
By analogy to the previous section, the solution can also be represented as the Whittaker function
\begin{equation}
P(r,\tau) = \exp \bigg\{-\frac{\alpha z}{2\sqrt{\alpha^2+2\beta^2}}\bigg\} z^{-\gamma/\beta^2} Z(z,\tau), \quad Z(z,\tau) \fallingdotseq G(z,\eta), \quad z= \frac{2r\sqrt{\alpha^2+2\beta^2}}{\beta^2}.
\end{equation}
\begin{equation}
    \left\{ {\begin{array}{l}
        G''+G\bigg(-1/4+\tilde{\lambda}/z+(1/4-\mu^2)/z^2 \bigg) =-\sigma z^{\gamma/\beta^2-1} e^{\alpha \sigma z/2},\\
        e^{-\alpha \sigma z/2} z^{-\gamma/\beta^2} G \rightarrow  \sigma/\eta,  \quad z\rightarrow 0, \\
        e^{-\alpha \sigma z/2} z^{-\gamma/\beta^2} G \rightarrow 0, \quad z\rightarrow +\infty.
    \end{array}} \right.
\end{equation}
where constants are denoted as
\begin{equation}
\sigma=1/\sqrt{\alpha^2+2\beta^2},
\quad \mu=\gamma/\beta^2-1/2, \quad \tilde{\lambda}=-\sigma(\alpha\gamma/\beta^2+\eta).
\end{equation}
The solution of nonhomogeneous boundary problem is
\begin{equation}
\begin{split}
G(z,\eta)= \sigma &\frac{\Gamma(1/2+\mu-\tilde{\lambda})}{\Gamma(1+2\mu)} \bigg( M_{\tilde{\lambda},\mu} (z) \int\limits_{0}^{z} \phi^{\gamma/\beta^2-1} e^{\alpha \sigma \phi/2} W_{\tilde{\lambda},\mu}(\phi)d\phi + \\
+ &W_{\tilde{\lambda},\mu} (z) \int\limits_{z}^{+\infty} \phi^{\gamma/\beta^2-1} e^{\alpha \sigma \phi/2} M_{\tilde{\lambda},\mu}(\phi)d\phi \bigg).
\end{split}
\end{equation}
The following transforms are similar to the geometric case (the formulas (57) and (59) are used)
\begin{equation}
\begin{split}
G(z,\eta)= \sigma \sqrt{z} \int\limits_{0}^{+\infty}\int\limits_{0}^{+\infty} &\phi^{\gamma/\beta^2 -1/2} \exp \bigg\{-\frac{z\cosh{\zeta}}{2}-\frac{\phi(\cosh{\zeta}-\alpha\sigma)}{2} \bigg\} \times \\
 &\times \coth^{2\tilde{\lambda}} \bigg ( \frac{\zeta}{2}\bigg) I_{2\mu} (\sqrt{z\phi} \sinh{\zeta})  d\phi d\zeta,
\end{split}
\end{equation}
\begin{equation}
\begin{split}
G(z,\eta) = 2\sigma& \int\limits_{0}^{+\infty} \exp\bigg\{ -\frac{z\cosh{\zeta}}{2}+\frac{z\sinh^2{\zeta}}{4(\cosh{\zeta}-\alpha \sigma)} \bigg\} \bigg( \frac{\cosh{\zeta}-\alpha \sigma}{2}\bigg)^{-\gamma/\beta^2} \times \\
&\times \sinh^{-1}(\zeta)\coth^{2\tilde{\lambda}} \bigg(\frac{\zeta}{2}\bigg) M_{-\gamma/\beta^2,\mu} \bigg( \frac{z\sinh^2{\zeta}}{2(\cosh{\zeta}-\alpha\sigma)} \bigg)d\zeta.
\end{split}
\end{equation}
The Whittaker function \begin{math} M_{\lambda,\mu} \end{math} has a simple special case \begin{math} M_{k,-k-1/2} (z) = e^{z/2} z^{-k} \end{math} (see Appendix C).
This gives
\begin{equation}
G(z,\eta) = 2 \sigma z^{\gamma/\beta^2} e^{\alpha\sigma z/2} \int\limits_{0}^{+\infty} \exp\bigg\{\frac{(\alpha^2\sigma^2-1)z}{2(\cosh{\zeta}-\alpha\sigma)} \bigg\}\sinh^{-1}(\zeta)\coth^{2\tilde{\lambda}} \bigg(\frac{\zeta}{2}\bigg) \bigg( \frac{\sinh{\zeta}}{\cosh\zeta-\alpha\sigma}\bigg)^{2\gamma/\beta^2} d\zeta.
\end{equation}
The change of variable also gives the Delta function representation
\begin{equation}
\ln{\tanh^2 \bigg(\frac{\zeta}{2}\bigg)}=\psi, \quad \frac{2d\zeta}{\sinh\zeta} =d\psi,
\quad \sinh\zeta=\frac{2e^{\psi/2}}{1-e^{\psi}}, \quad \cosh\zeta -\alpha\sigma = \frac{(1+\alpha\sigma)e^{\psi}-\alpha\sigma+1}{1-e^{\psi}}.
\end{equation}
Then the bond value is equal to (\begin{math} T-t=\tau \end{math})
\begin{equation}
 P(r,t)=\bigg( \frac{2 e^{-\tau/2\sigma}}{(1+\alpha\sigma)e^{-\tau/\sigma} -\alpha\sigma+1}\bigg)^{2\gamma/\beta^2} \exp\bigg\{-\frac{\alpha \tau}{\beta^2}+
 \frac{(\alpha^2\sigma^2-1)(1-e^{-\tau/\sigma})r}{\beta^2\sigma\big((1+\alpha\sigma)e^{-\tau/\sigma} -\alpha\sigma+1 \big)} \bigg\}  .
\end{equation}

\subsection{Laguerre polynomials expansions}
Another approach in this problem is the Laguerre polynomials expansions. The solution original PDE problem (68) can be find in series representation
\begin{equation}
    P(r,t)=\sum_{k=0}^{+\infty} C_k G_k(r) e^{-\lambda_k \tau}, \quad (\beta^2 r /2)G''_k+(\alpha r +\gamma) G'_{k}-rG=0.
\end{equation}
the last equation can be reduced to the equation
\begin{equation}
xu''+(\alpha^{*}-x+1)u'+nu=0
\end{equation}

A solution of it is the generalized Laguerre polynomials \begin{math} L_n^{\alpha^{*}}(x) \end{math} (see Appendix C). Using boundary conditions and orthogonality of generalized Laguerre polynomials constants \begin{math} C_k \end{math} can be determined. The resulting series is the characteristic function of Laguerre polynomials (see Appendix C). This approach is considered in \cite{BUT}.

\subsection{The double Laplace transform method}
The original PDE problem can be reduced to the first-order ODE. For this the Laplace transform under spatial variable \begin{math} r \end{math} and time variable is used.
\begin{equation}
\begin{split}
    &P(r, \tau) \fallingdotseq Q(q,\tau), \quad
    P_{r} \fallingdotseq qQ(q,\tau)-P(0,\tau), \quad
    rP(q, \tau) \fallingdotseq -Q_{q}(q,\tau), \quad\\
    rP_{r} \fallingdotseq &-qQ_{q}(q,\tau)-Q(q,\tau),\quad
    rP_{rr} \fallingdotseq -q^2 Q_{q}(q,\tau)-2q Q(q,\tau), \quad Q(q,\tau) \fallingdotseq W(q,\eta).
\end{split}
\end{equation}
This reduces our problem to the first order PDE.
\begin{equation}
    \left\{ {\begin{array}{l}
        (\gamma^2 q^2 /2 -\beta q +1) W_{q} + (\gamma^2 q - \alpha q +\beta+\eta)W  =1-\alpha, \\
        Q(0)=0, \\
        Q(+\infty)=1/\eta. \\
    \end{array}} \right.
    \end{equation}
But the inversion procedure is very complicated due to these reasons: firstly the image of double Laplace transform is the function of two complex variables, and secondly this function has a structure of multiplication of incomplete Beta and Gamma special functions. However in case of integer \begin{math} \eta \end{math} (i.e. discrete spectrum) the simplifications are essential: these special functions are reduced to the polynomials. This leads to the Laplace transform under spatial variable in series representation (79). It also reduces original problem to the infinite system of first order ODE. Let me mention that the Laplace transform under spatial variable in equation (87) is a one of approaches of determination its Laguerre polynomials structure.

\section{About one modification of Constantinides - Ingersoll model}
This model also fits on the Whittaker functions framework
    \begin{equation}
    dr(t)= \alpha r^2 dt + \beta r dW_t.
    \end{equation}
The corresponded boundary problem is
    \begin{equation}
    \left\{ {\begin{array}{l}
        (\beta^2 r^2/2) P_{rr}+\alpha r^2 P_{r}-rP =P_{\tau}, \\
        P(r,0)=1, \\
        P(0,\tau)=1, \\
        P(+\infty,\tau)=0. \\
    \end{array}} \right.
    \end{equation}
And the solution of it:
\begin{equation}
\begin{split}
&\quad \quad P(r,t)=\Gamma(1+1/\alpha) e^{-\alpha r /\beta^2} W_{-1/\alpha,1/2} (2\alpha r /\beta^2) +\frac{2\sqrt{2} \sqrt{r}}{\sqrt{\alpha} \beta \sqrt{\pi} \pi} \times \\
\times \int\limits_{1}^{+\infty}\int\limits_{1}^{+\infty} &\frac{(\theta^2-1)^{1/\alpha-1} \sinh(\pi\sqrt{\xi^2-1}/2) e^{-\alpha r \theta^2 /\beta^2 -\beta^2 (T-t) \xi^2 /8} K_{i\sqrt{\xi^2-1}/2}(\alpha r \theta^2 /\beta^2)}{\xi \theta^{2/\alpha}} d\xi d\theta.
\end{split}
\end{equation}

\appendix
\section{Solution of hypergeometric equation}
\par
According to \cite{NU} the hypergeometric equation is defined as
\begin{equation}
    \sigma(x) y'' + \tau(x) y' +\omega y =0
\end{equation}
where \begin{math} \sigma (x) \end{math} is the second order polynomial and \begin{math} \tau(x) \end{math} is the first order polynomial. Any solution of this equation can be represented as the integral on the complex plane
\begin{equation}
    y(x)=\frac{C_\nu}{\rho(x)}  \int\limits_{C}{} \frac{\sigma^\nu(s) \rho(s) }{(s-x)^{\nu+1} }  ds
\end{equation}
where \begin{math} C_\nu \end{math} is a constant, \begin{math} \rho(x) \end{math} and \begin{math} \nu \end{math} satisfy the following conditions
\begin{equation}
   (\sigma \rho)' = \tau \rho , \quad  \omega = -\nu \tau' - \frac {\nu (\nu-1)}{4} \sigma'' .
\end{equation}

The path of integration \begin{math} С \end{math} is set with these restrictions
\begin{equation}
    \frac{\sigma^{\nu+1}(s) \rho(s)}{(s-x)^{\nu+2}} \bigg|^{s_2} _ {s_1} =0
\end{equation}
where \begin{math} s_1 \end{math} and \begin{math} s_2 \end{math} are the ends of the path \begin{math} С \end{math} and
\begin{equation}
    \frac {d^k}{dx^k} \bigg[ \int\limits_{C}{} \frac{\sigma^\nu(s) \rho(s) }{(s-x)^{\nu+1} }  ds \bigg ] = (\nu+1){...} (\nu+k) \int\limits_{C}{} \frac{\sigma^\nu(s) \rho(s) }{(s-x)^{\nu+1} }  ds , \quad k=1,2.
\end{equation}
This result generalizes Rodrigues formula \cite{NU} for hypergeometric ODE solution with integer \begin{math} \omega \end{math} (i.e. Orthogonal polynomials).
\par

Application of this formula in equation (15) is given by these formulas:
\begin{equation}
    \sigma(x) = \frac{\beta^2 x^2}{2}, \quad \tau(x)=\alpha x - 1, \quad \omega = -1.
\end{equation}
The solutions of ODE for \begin{math} \rho \end{math} and algebraic equation for \begin{math} \nu \end{math} are
\begin{equation}
    \frac{\rho^{'}}{\rho} = \frac{2(\alpha-\beta^2)}{\beta^2 x} - \frac{2}{\beta^2 x^2} \Longleftrightarrow
    \rho(x) = x^{{(2\alpha-2\beta^2)}/{\beta^2}}  e^{2/\beta^2 x},
\end{equation}
\begin{equation}
    \frac{\beta^2}{2} \nu^2 - \bigg( \frac{\beta^2}{2} - \alpha \bigg) \nu -1=0 \Longleftrightarrow
    \nu_{1,2}=  \bigg(\frac{1}{2}  -\frac{\alpha}{\beta^2} \bigg )
    \pm \sqrt{\frac{1}{4} + \frac{2}{\beta^2} -\frac{\alpha}{\beta^2} +\frac{\alpha^2}{\beta^4}}.
\end{equation}
Ends of the integration path are defined as \begin{math} s_0=-\infty \end{math} and \begin{math} s_1=0 \end{math}. Then the solution of equation (15) can be represented as this infinite integral
\begin{equation}
    F(x)=x^{2-2\alpha/\beta^2}e^{-2/{\beta^2 x}}\int\limits_{-\infty}^{0} s^{2\nu+ 2\alpha/\beta^2 -2} e^{2/{\beta^2 s}} (s-x)^{-(1+\nu)} ds
\end{equation}
or
\begin{equation}
    F(x)=x^{2-2\alpha/\beta^2}e^{-2/{\beta^2 x}}\int\limits_{0}^{+\infty} s^{2\nu+ 2\alpha/\beta^2 -2} e^{-2/{\beta^2 s}} (s+x)^{-(1+\nu)} ds.
\end{equation}
This integral is well-known \cite{GR}
\begin{equation}
\begin{split}
    &\qquad \int\limits_{0}^{+\infty} s^{2\nu+ 2\alpha/\beta^2 -2} e^{-2/{\beta^2 s}} (s+x)^{-(1+\nu)} ds = \\
    = \bigg ({\frac{2}{\beta^2}} \bigg )^{\nu+{\alpha/\beta^2}-1} & x^{(-1+\alpha / \beta^2)} \Gamma (2-\nu- 2\alpha / \beta^2) e^{1/{\beta^2 x } } W_{-1+\alpha / \beta^2, -\nu -\alpha/\beta^2 +1/2 } \bigg(\frac {2} {\beta^2 x} \bigg).
\end{split}
\end{equation}
Basing on linear relations between \begin{math} M_{\lambda,-\mu} \end{math}, \begin{math} M_{\lambda,\mu} \end{math} and  \begin{math} W_{\lambda,\mu} \end{math} a solution of (15) can be defined as:
\begin{equation}
    F(x)= x^{1-\alpha/\beta^2}e^{-1/{\beta^2 x}} M_{(-1+\alpha / \beta^2), -\nu -\alpha/\beta^2 +1/2 } \bigg(\frac {2} {\beta^2 x} \bigg) = x^{-\lambda}e^{-1/{\beta^2 x}} M_{\lambda, \pm \mu} \bigg(\frac {2} {\beta^2 x} \bigg).
\end{equation}

\section{Time-dependent coefficients models: The Ho-Lee and Hull-White model}
Ho-Lee model and Hull White models are introduced in 1986 and 1988 respectively and contrary to all previous models have time-dependent coefficients. Consider Ho-Lee model:
    \begin{equation}
    dr(t)=\alpha(t) dt + \gamma(t) dW_t.
    \end{equation}
The corresponded boundary problem has time-dependent coefficients
    \begin{equation}
    \left\{ {\begin{array}{l}
        \gamma^2(\tau) P_{rr} + \alpha(\tau)P_r -rP =P\tau, \\
        P(r,0)=1, \\
        P(0,\tau)=1, \\
        P(+\infty,\tau)=0. \\
    \end{array}} \right.
    \end{equation}
The Laplace transform is made in the manner of subsection 5.3 and it gives also the first order PDE:
\begin{equation}
    \left\{ {\begin{array}{l}
        Q_{q}-Q_{\tau} +Q(\gamma^2(\tau)q^2 +\alpha(\tau)q) =-\alpha(\tau), \\
        Q(q,0)=1, \\
        Q(0,\tau)=0, \\
        Q(+\infty,\tau)=1. \\
    \end{array}} \right.
    \end{equation}
The structure of solution can be represented by \begin{math} Q(q,\tau)=e^{g(q,\tau)} \end{math}. The homogeneous equation for \begin{math} g \end{math} is
\begin{equation}
    g_{q}-g_{\tau} = \gamma^2(\tau)q^2 +\alpha(\tau)q.
\end{equation}
For solving this problem another Laplace transform  \begin{math} g(q,\tau) \fallingdotseq H(h,\tau) \end{math} is used. Equation (C9) transforms to
\begin{equation}
    H_{\tau}-hH = 2\gamma^2(\tau)/h^2 +\alpha(\tau)/h.
\end{equation}
After that the solution of nonhomogeneous PDE is found in representation \begin{math} Q=C(q,\tau)e^{g(q,\tau)} \end{math} (This is analogical to the ODE problems). Function \begin{math} C(q,\tau) \end{math} is a solution of equation that also can be solved by Laplace transform technique.
\begin{equation}
C_{q}-C_{\tau} = -\alpha(\tau) /e^{g(q,\tau)}.
\end{equation}
In Hull - White model
    \begin{equation}
    dr(t)=\big(\alpha(t)+\beta(t)r(t)\big) dt +\gamma(t) dW_t.
    \end{equation}
all steps of the method are fully analogical. However the corresponding first order PDE is more formidable, but there are no any principal problems in this scheme.

\section{Necessary facts of Bessel and Whittaker functions}
\subsection{Bessel functions}
\subsubsection{Definition}
The Bessel function \begin{math} J_{\nu}(z) \end{math} and modified Bessel function \begin{math} I_{\nu}(z) \end{math}  can be defined as a solution of this ODE`s (This ODE cames form Helmholtz equation \begin{math} \Delta U(x,y) +\lambda^2U(x,y)=0 \end{math} in the radius/angle decomposition):
\begin{equation*}
Z''+\frac{1}{z}Z'+\bigg(1-\frac{\nu^2}{z^2} \bigg)Z=0, \quad Z''+\frac{1}{z}Z'+\bigg(1-\frac{\nu^2}{z^2} \bigg)Z=0.
\end{equation*}
The other liner independent solution of these equation is the Neumann function \begin{math} N_{\nu}(z) \end{math} and McDonald function \begin{math} K_{\nu}(z) \end{math} respectively. The relations between these functions are given by these formulas:
\begin{equation*}
 N_{\nu}(z)= \frac{1}{\sin{\nu\pi}} \big[\cos{\nu\pi} J_{\nu}(z)- J_{-\nu}(z) \big], \quad
 K_{\nu}(z)= \frac{\pi}{2\sin{\nu\pi}} \big[I_{-\nu}(z)- I_{\nu}(z) \big], \quad \nu \neq Z.
\end{equation*}
\begin{equation*}
\begin{split}
\quad I_{\nu}(z)= e^{-\pi\nu i/2} J_{\nu}(e^{\pi i/2}z), \quad [-\pi/2 < \arg z <= \pi/2] , \\
\quad J_{\nu}(z)= e^{3\pi\nu i/2} I_{\nu}(e^{-3\pi i/2}z), \quad [\pi/2 < \arg z <= \pi].
\end{split}
\end{equation*}

\subsubsection{Series and integral representations}
\begin{equation*}
\begin{split}
&1. \quad J_{\nu}(z)=\sum_{k=0}^{+\infty} \frac{(-1)^k}{k!\Gamma(\nu+k+1)} \bigg(\frac{z}{2}\bigg)^{\nu+2k}, \quad I_{\nu}(z)=\sum_{k=0}^{+\infty} \frac{1}{k!\Gamma(\nu+k+1)} \bigg(\frac{z}{2}\bigg)^{\nu+2k}, \\
&2. \quad J_{\nu}(z)=\frac{2}{\pi} \int\limits_{0}^{+\infty} \sin\bigg( z \cosh{t} -\frac{\nu\pi}{2} \bigg)\cosh{\nu t} dt, \quad K_{\nu}(z)=\int\limits_{0}^{+\infty} e^{-z\cosh{t}}\cosh{\nu t} dt.
\end{split}
\end{equation*}
\subsubsection{Wronskians}
\begin{equation*}
Wronskian[J_{\nu}(z), N_{\nu}(z)] = \frac{2}{\pi z}, \quad Wronskian[I_{\nu}(z), K_{\nu}(z)] = -\frac{1}{z}.
\end{equation*}
\subsubsection{Asymptotic expansions}
\begin{equation*}
\begin{split}
    &1. \quad K_{\nu} (z) \thicksim \frac{\pi 2^{\nu-1}}{\sin{\nu\pi} \Gamma(1-\nu)z^{\nu}}, \quad I_{\nu} (z) \thicksim \frac{z^{\nu}}{2^{2\nu}\Gamma(1+\nu)}, \quad z \rightarrow 0, \\
    &2. \quad K_{\nu} (z) \thicksim \sqrt{\frac{\pi}{2z}}e^{-z}, \quad I_{\nu} (z) \thicksim \sqrt{\frac{1}{2\pi z}}e^{z}, \quad z \rightarrow +\infty.
\end{split}
\end{equation*}
\subsubsection{Analytic continuations}
\begin{equation*}
\begin{split}
&1. \quad J_{\nu} (e^{m\pi i}z)=e^{m \nu\pi i}J_{\nu} (z), \quad I_{\nu} (e^{m\pi i}z)=e^{m \nu\pi i}I_{\nu} (z), \\
&2. \quad K_{\nu} (z)=e^{-m\nu \pi i }K_{\nu } (z) -i\pi \frac{\sin m\nu \pi}{\sin \nu\pi} I_{\nu} (z), \quad \nu \neq Z.
\end{split}
\end{equation*}

\subsection{Whittaker functions}
\subsubsection{Definition}
The Whittaker functions \begin{math} M_{\lambda,\mu}(z),\quad M_{\lambda,-\mu}(z), \quad W_{\lambda,\mu}(z), \quad W_{-\lambda,\mu}(-z), \end{math} are the solutions of this ODE
\begin{equation*}
G''+\bigg( -\frac{1}{4}+\frac{\lambda}{z}+ \frac{1/4-\mu^2}{z^2} \bigg)=0.
\end{equation*}
These functions have relation between them and confluent hypergeometric functions \begin{math} \Phi (\alpha,\beta,\gamma) \end{math}:
\begin{equation*}
\begin{split}
&1.\quad M_{\lambda,\mu}(z) = z^{\mu+1/2} e^{z/2} \Phi(\mu-\lambda+1/2,2\mu+1,z),\\
&2. \quad W_{\lambda, \mu} (z)= \frac{\Gamma(-2\mu)}{\Gamma(1/2-\mu-\lambda)} M_{\lambda,\mu}(z)+ \frac{\Gamma(2\mu)}{\Gamma(1/2+\mu-\lambda)} M_{\lambda,-\mu}(z).
\end{split}
\end{equation*}
\subsubsection{Integral representation}

\begin{equation*}
\begin{split}
&1.\quad M_{\lambda,\mu}(z) =\frac{z^{\mu+1/2}} {2^{2\mu} B(\mu+\lambda+1/2, \mu-\lambda+1/2)}
\int\limits_{-1}^{1} (1+t)^{\mu-\lambda-1/2} (1+t)^{\mu + \lambda-1/2} e^{zt/2} dt, \\
&2. \quad W_{\lambda,\mu}(z) =\frac{z^{\mu+1/2}e^{-z/2}}{\Gamma(\mu-\lambda+1/2)}
\int\limits_{0}^{+\infty} t^{\mu-\lambda-1/2} e^{-t} (1+t/z)^{\mu + \lambda-1/2}dt.
\end{split}
\end{equation*}
where \begin{math} B(x,y) \end{math} is the Euler beta function.

\subsubsection{Wronskians}
\begin{equation*}
Wronskian \big[ M_{\lambda,\mu}(z), M_{\lambda,-\mu}(z) \big] =-2\mu ,
\quad Wronskian \big[ M_{\lambda,\mu}(z), W_{\lambda,\mu}(z) \big] =-\frac{\Gamma(1+2\mu)}{\Gamma(1/2+\mu-\lambda}.
\end{equation*}

\subsubsection{Asymptotic}
\begin{equation*}
\begin{split}
  &1.\quad  M_{\lambda,\mu} (z) \thicksim z^{\mu+1/2}(1+O(z)), \\
  &2.\quad W_{\lambda, \mu} (z)= \frac{\Gamma(2\mu)}{\Gamma(1/2+\mu-\lambda)} z^{1/2-\mu}+ \frac{\Gamma(-2\mu)}{\Gamma(1/2+\mu-\lambda)} z^{1/2+\mu} +O\bigg( z^{3/2-\Re\mu}\bigg)
, \quad z \rightarrow 0, \\
  &3. \quad  M_{\lambda,\mu} (z) \thicksim \frac{\Gamma(1+2\mu)}{\Gamma (1/2 + \mu - \lambda)}z^{-\lambda} e^{z/2} ,
    \quad W_{\lambda, \mu} (z) \thicksim e^{-z/2} z^{\lambda},
    \quad z \rightarrow +\infty.
\end{split}
\end{equation*}
\subsubsection{Relation to the other functions}
\begin{equation*}
\begin{split}
&1. \quad W_{n+\mu+1/2,\mu} (z) =(-1)^{n} (2\mu+1)_{n} M_{n+\mu+1/2,\mu} (z) = (-1)^{n} n! z^{\mu+1/2} e^{-z/2} L_{n}^{2\mu}(z),
\\
&2.\quad M_{\lambda,\lambda-1/2}(z) = W_{\lambda,\lambda-1/2}(z) = W_{\lambda,-\lambda-1/2}(z) =e^{-z/2} z^{\lambda}, \quad M_{\lambda,-\lambda-1/2}(z) =e^{z/2} z^{-\lambda}.
\\
&3. \quad M_{0,\mu} (z) = 2^{2\mu} \Gamma(\mu+1) \sqrt{z} I_{\mu}\bigg( \frac{z}{2}\bigg), \quad W_{0,\mu}(z) = \sqrt{\frac{z}{\pi}} K_{\mu} \bigg( \frac{z}{2}\bigg),
\\
&4. \quad M_{-1/4,1/4}(z^2) =\frac{1}{2} e^{z^2/2} \sqrt{\pi z} erf(z), \quad W_{-1/4,\pm1/4}(z^2) =e^{z^2/2} \sqrt{\pi z} erfc(z) ,
\\
&5. \quad W_{1/4+p/2,-1/4}(z^{2}/2) =2^{-1/4-p/2}D_{p}(z).
\end{split}
\end{equation*}
where \begin{math} L_{n}^{\mu} \end{math} is generalized Laguerre polynomials,
\begin{math} erf(z) \end{math} and \begin{math}  erfc(z) \end{math} are Error functions, and \begin{math}  D_{p} (z) \end{math} is Parabolic cylinder function.
\subsubsection{Laguerre polynomials}
\begin{equation*}
L_n^{\alpha^{*}}(x) = \frac{1}{n!} x^{-\alpha^{*}} \frac{d^n}{dx^n}(e^{-x}x^{n+\alpha^{*}}),
\quad \sum_{n=0}^{+\infty} L_n^{\alpha^{*}}(x) z^{n} = (1-z)^{-\alpha-1} \exp\bigg\{\frac{xz}{z-1}\bigg\}, \quad |z|<1.
\end{equation*}.

\end{document}